# The hydrogen molecule in a vice

Matthias Stocker, Simon Röger, and Berndt Koslowski

*Institute of Solid State Physics, University of Ulm, Albert-Einstein-Allee 11, D-89069 Ulm*

We conducted experiments on the role of molecular hydrogen physisorbed between two metallic surfaces. Such hydrogen leads to strong signatures in inelastic electron tunneling spectroscopy exceeding the amplitude of typical inelastic transitions by an order of magnitude and is accompanied by a step in the tunneling current leading under certain circumstances to a huge negative differential conductance. We show that the molecular vibration opens an inelastic channel as expected but simultaneously stalls the total elastic channel due to the forces exerted by the vibrating molecule on the electrodes. The required compliance of the stylus is surprisingly large but is possible in the wide spectrum of experimental conditions. Additionally, the inelastic channel experiences a saturation from which the lifetime of the molecular vibration can be deduced to be approximately 1 ns. This experiment allows resolving the forces corresponding to specific vibrational states of a single molecule.

Hydrogen, the simplest atom, the simplest molecule, the most abundant element in the universe, imposes challenges to science and technology at all times. Apart from the fundamental role of hydrogen in chemistry and biology, in physics, just the phase diagram of the hydrogen is most probably the most chameleonic of all materials[1,2,3,4]: Hydrogen may be a molecular gas, a plasma, an atomic fluid, a molecular fluid, a superfluid, a superconductor, a metallic fluid or a metallic solid thus being thought to be responsible for the largest object in our solar system beyond the sun, the magnetosphere of Jupiter[5,6]. Also, in solid state physics and surface science, hydrogen is a pervasive companion. As it easily diffuses through and over all materials, it is the only supposable evil left even at low temperatures and in ultra-high vacuum.

A decade ago, Gupta et al. reported that physisorbed molecular hydrogen leads to marked signatures in scanning tunneling spectroscopy (STS)[7]. This was surprising since physisorbed hydrogen molecules – with no electronic states close to the Fermi level and a very weak interaction with the surface – would be supposed to be unobtrusive in STM. Later, Weiss at al. showed that the hydrogen molecules have a dramatic influence on the imaging mechanism in scanning tunneling microscopy (STM) by increasing the spatial resolution down to faint sub-molecular structures[8]. By a combined STM and atomic force microscopy (AFM) study, Lotze at al. recovered that hydrogen in the tunneling junction may impel energy to the lever oscillation via a stochastic resonance[9]. Li et al.[10] and Natterer et al.[11] observed similar structures in inelastic electron tunneling spectroscopy (IETS) but focused on the excitation of molecular rotations revealing the nuclear spin of the molecules. Halbritter et al.[12] calculated the *I-V* curves based on a two-level system with two distinct conductivity values. They aimed at junctions at the high-conductivity end close to the quantum conductance unit[13] but utilized observations made already by Gupta et al.: in the transition region random pulse trains appeared with two distinct current values. A physical description of the two-level system, however, leading conclusively to the observed signatures is still lacking.

We will show in this contribution that the observed phenomena are the manifestation of forces exerted by the confining walls to keep a single vibrating molecule in its position. The experiment allows controlling the vibrational states of the molecule and determination of the related forces.

Samples are Au single-crystals exposing the (111) face at a miss-cut angle <0.1°. The surface has been prepared by sputter-anneal cycles[14] until the surface showed the clean 22×√3 reconstruction with no adsorbates at the elbow sites and straight steps separating the terraces. $H_2$ has been dosed by floating the chamber with the gas to a pressure of $1\times10^{-6}$ mbar via a leak valve while running *I-V* scans with the STM at the Au(111). The dosing was stopped after about 10 minutes when the *I-V* scans showed clearly the signatures of the hydrogen molecules. Tunneling tips were prepared from Ag wires in a two-phase etching process similar to that proposed in ref. 15. The tips were cleaned in situ by annealing to 700 °C and careful conditioned by field emission and desorption against a Nb(110) surface.

Partly, the tips were sputtered using Ar ions. The STM is a custom-built microscope operated at a base temperature of 5 K in ultra-high vacuum ($10^{-10}$ mbar)[16].

Figure 1 shows $\partial_V I$-$V$ (a) and $I$-$V$ curves (b) taken at different tip-sample separations. The most significant features in $\partial_V I$-$V$ curves are the dominant negative peak which shifts weakly with varying tip-sample separation, and a negative step of the conductivity occurring across that peak. These signatures are very strong – the peak may go negative (negative differential conductivity) and the step may amount up to ~70% of the conductivity below the peak – such that the measurement of the second derivative would be difficult due to the high dynamic range required. Far away from the sample (see supplementary material) peak and step may change sign.

In order to gain insight into the physics behind such $\partial_V I$-$V$ curves, we analyzed the phase integral from $I$-$z$ measurements shown in Figure 2(a). In the Wentzel-Kramers-Brillouin (WKB) approximation[17], the transmission probability function is assumed to be $T = e^{-P}$ with the phase integral $P = \int_{z_1}^{z} k(z')dz' = \frac{1}{\hbar}\int_{z_1}^{z}\sqrt{2m(\Phi - E)}dz'$ between the classical turning points $z_1$ and $z$ with the potential in the tunneling barrier, $\Phi$, the energy of the electron, $E$, and the mass of the electron, $m$. Assuming the density of states of tip and sample being constant in the relevant energy range, the tunneling current is then $I \approx \kappa \int_0^{eV} T\, dE \approx eV\kappa T$ with $\kappa$ a constant which we set to unity for simplicity, the electron charge, $e$, and the applied bias, $V$. Hence $\frac{\partial_z I}{I} = -\partial_z P = -k(z)$. We measured a set of $I$-$z$ curves at different biases (4 out of 20 in such a sequence are shown in Figure 2(a)) and calculated the derivative and the normalization numerically. Accordingly, we obtain a wave number which shows a Lorentzian-like peak having an amplitude corresponding to the background of $k_0 = 20..23$ nm$^{-1}$ which is expected for the barrier height between Ag and Au ($\Phi \cong 4..5$ eV). The Lorentzian depends on the applied bias as the center of the peak, $z_c$, shifts towards the sample. The width of the peak is approximately 40 to 50 pm and is independent of the bias.

With the above behavior of $k$ at hand, we can calculate easily the characteristic features of measured $\partial_V I$-$V$ curves. Assuming that $k$ is indeed a Lorentzian, we set

$$k(z) = k_0 + \frac{k_1}{1 + 4\left(\frac{z - z_c}{w_z}\right)^2} \qquad (1)$$

with the width of the Lorentzian, $w_z$. Then, the phase integral becomes

$$P = k_0 z + \frac{k_1 w_z}{2}\left(\arctan\left(2\frac{z - z_c}{w_z}\right) + \arctan\left(2\frac{z_c}{w_z}\right)\right). \qquad (2)$$

If now $z_c$ depends on bias, differentiation of the tunneling current with respect to bias generates terms in the prefactor containing the Lorentzian again reading

$$\partial_V I = e\left(1 - V \cdot k_1 \cdot \partial_V z_c \left(\frac{1}{1 + 4\left(\frac{z_c}{w_z}\right)^2} - \frac{1}{1 + 4\left(\frac{z - z_c}{w_z}\right)^2}\right)\right) \times e^{-P} \qquad (3)$$

and normalized with $I$

$$\frac{\partial_V I}{I} = \frac{1}{V} - k_1 \cdot \partial_V z_c \left(\frac{1}{1 + 4\left(\frac{z_c}{w_z}\right)^2} - \frac{1}{1 + 4\left(\frac{z - z_c}{w_z}\right)^2}\right). \qquad (4)$$

Thus, a $\partial_V I$-$V$ curve comprises a smeared step as an arc tangent in the exponential of $P$, and a Lorentzian peak centered at $z_c$ of width $w_z$, which transforms to a center and width in $V$, $V_c$ and $w_V \approx |w_z / \partial_V z_c|$, respectively. In principle one expects two transitions in the $\partial_V I$-$V$ curve, but one is, as will be seen later, negligible ($z_c / w_z \gg 1$) and physically dispensable. For the same arguments, the step in the exponential can be approximated by a simple step function. An exemplary fit to measured $\partial_V I$-$V$ data is shown in Figure 2(b) where we replaced the Lorentzian peak by a Gaussian and the exponential of the arc tangent by an error function because the measurement is obviously subject to inhomogeneous broadening at a temperature of 6.5 K[18]. The correlation of measured data and the model function is typically above 99.6%. In Figure 2(c) we show the normalized conductivity according to equation (4) for different tip-sample separations after the $1/V$ term has been removed. One recognizes a peak shifting with separation, and the amplitude of the peak reduces by about 35% for a change of the separation by about 0.1 nm.

With the accurate description of experimental data we are able to analyze the dependence of the characteristic parameters on the tip-sample separation. The characteristic parameters are the center of the peak of the conductivity, $V_c$, and optionally its width, $w_V$. The other characteristic parameters are the current step occurring at $V_c$ with respect to the current corresponding to the low bias conductivity at $V_c$, $\alpha_{elastic}$ (cfg. 1(b)), and the efficiency of the inelastic channel, $\eta_{inelastic}$, with respect to the static conductivity at $V_c$ including the current step. In other words, we interpret data such that the elastic channel switches conductivity at $V_c$ and we measure the efficiency of the inelastic channel with respect to the elastic conductivity above $V_c$. Note, that, from a mathematical point of view, the change of the elastic conductivity is due to the fact that the Lorentzian peak in $k$ shifts into the tunneling barrier by the bias dependence of the peak center. That peak resides outside the barrier if $V < V_c$ and inside the barrier if $V > V_c$ such that the damping of the elastic channel is due to the change of the phase integral across that peak, and, hence,

$$1 + \alpha_{elastic} = e^{-\Delta P} = e^{-k_1 w_z \frac{\pi}{2}} \rightarrow -\alpha_{elastic} \approx \Delta P \qquad (5)$$

if $\alpha_{elastic}$ is small enough.

In Figure 3 we show the characteristic parameters of those measurements which showed only a single peak in the conductivity. We consider this being evidence of a single hydrogen molecule in the tunneling junction. In the particular measurements shown in Figure 3(a), $\eta_{inelastic}$ is positive and fades off close to the sample. It is important to mention that, due to the huge current step occurring at $V_c$, one has to change measurement and evaluation techniques such that one looses quite a lot of resolution with respect to the limits achieved in IETS earlier[19]. In the particular measurements shown, the conductivity change of the elastic channel, $\alpha_{elastic}$, cfg. 3(b), is between -50 % to -70 % close to the sample and drops quickly with increasing separation. Small positive values have been observed for smaller currents (larger $z$). Figure 3(c) shows the dependence of the center of the peaks/steps in $\partial_V I$-$V$ curves for different samples. The absolute values for different samples/tips are different which indicates that the resonance depends on individual properties of the tunneling junction. The resonance shifts down systematically for increasing tip-sample separation which is indicative of a weaker binding of the $H_2$ molecule for increasing separation. The negative sign of the slope of $V_c(z)$ is consistent with the observation that $z_c$ shifts towards or into the barrier with increasing bias. Consequently, one finds a negative conductance step close to the sample.

In agreement with earlier discussions, we attribute the observed phenomenon to the vibration of the hydrogen molecule in the physisorption well at the sample surface. The physisorption of hydrogen at metal surfaces has been studied decades ago and the physisorption parameters are well known for most noble or semi-noble metal surfaces[20,21]. A complication could arise in the presence of a curved probe. However, for a not too sharp probe the deviation should be small[22]. Therefore, we will confine ourselves to the superposition of two independent potential wells of flat surfaces leaving some space for energy shifts. Figure 4(a) shows the undistorted potentials of sample (Au(111), left side) and tip (Ag(111), right side) and its superposition for a tip-sample separation of 0.4 nm, 0.6 nm, and 0.8 nm. Note, that the difference between minima of the physisorption potential of the undistorted well and the superposition indicates the depth of the lateral potential well which confines the molecule to the center underneath the tip. At a separation greater than 0.6 nm a double well develops. For the current experiments we expect the range $z < 0.6$ nm to be relevant.

The energy levels of the vibrating $H_2$ molecule can be calculated numerically by applying Numerov's method to the physisorption potential[23,24]. Figure 4(b) shows the first ($v = 0 \to 1$, solid curve), the second ($v = 0 \to 2$, dotted curve) vibrational transition perpendicular to the surfaces, and the depth of the lateral well in dependence on the tip-sample separation. The energy of the $v = 0 \to 1$ transition is in the range 5-25 meV and decreases for increasing separation which is in accordance with experiment. It reaches a minimum at about 0.63 nm. As a consequence, $\alpha_{elastic}$ could change sign at larger separation as observed experimentally (compare supplementary material). The $v = 0 \rightarrow 2$ transition is close to the lateral depth of the well (Figure 4(b), dashed curve) indicating that the detection of that transition might be difficult because the molecule could be kicked out of the well by exciting that vibration. The spatial extent of the vibrational wave function of the hydrogen molecule is huge. Even the ground state extents over almost 0.2 nm perpendicular to the surface. The rotation of the molecule ($j = 0 \rightarrow 2$ transition for para-$H_2$) is expected at ~42 meV and is hindered by the dependence of the physisorption potential and the induced dipole moment on the rotation angle. In our measurements involving just one molecule/peak the rotational transition was always weak compared to the vibrational transition.

It is known that the efficiency of an inelastic channel may appear negative due to a negative retroaction of the inelastic excitation to the elastic channel[25]. The efficiency of the inelastic channel, however, is considered being positive leading to a positive change of the conductivity. Applying this notion to the present case, an electron may excite the vibration of the hydrogen molecule opening the inelastic channel of efficiency $\eta_{inelastic} > 0$, and, in turn, the vibrating molecule changes the tunneling probability function, $T$, until the excitation decays. Hence, a saturation is expected if the tunneling rate, $\Gamma_s$, times the inelastic efficiency exceeds the inverse lifetime of the vibration, $\tau_{vib}^{-1}$. Data shown in Figure 3 support that idea: the inelastic efficiency drops off continuously for increasing conductivity because the displayed range of $\sigma_0$ is entirely in saturation, i.e., in the regime of strong charge-vibron coupling. One obtains a tunneling rate of $\Gamma = \sigma_0 V_c/e$ corresponding to $\tau_{vib} \cong (\Gamma \eta_{inelastic})^{-1} \cong (1.7 \pm 0.3)$ ns (Figure 3(a) inset). The lifetime, $\tau_{vib}$, has been calculated and measured to be approximately 1 ps[26,27]. It increases quickly with increasing separation from the surface due to a weaker interaction with the metal electrons. The presence of the second electrode, however, may pull the equilibrium position of the molecule off the surface and, therefore, may lead to an enhanced lifetime.

The task left is to understand how the vibrating $H_2$ molecule affects the tunneling probability function. To find an *ansatz* we return to the phase integral, $P$. A positive contribution to $P$ of the observed magnitude would correspond to an extra path segment, $\delta s$, of $\Delta P = k_0 \delta s$ or $\delta s \approx -\alpha_{elastic}/k_0$. The crucial point is now to interpret this path segment as an elastic deformation of the tip sensor causing a change of the length of the tunneling junction equal to $\delta s$. If we take the ($v = 0 \to 1$) transition with energy $E_{0 \to 1}$ from Figure 4(b), $\partial_z E_{0 \to 1}$ corresponds to a change of the force acting on the electrodes due to the transition taking place. One observes a static force because the occupation probability of level 1 is approximately 1 for $|V| > V_c$ and 0 for $|V| < V_c$ disregarding the transition region close to $V_c$. Hence, the deformation is $\delta s = -C\, \partial_z E_{0 \to 1}$ with the compliance of the tip, $C$, if $\delta s$ is small enough that $\partial_z E_{0 \to 1}$ does not change significantly. Applying $k_1 = k_0$ to equation (2) one obtains $w_z = \delta s \cdot 2/\pi$ and, generalized to all transitions in *I-V* measurements with $z$ = const.,

$$\alpha_{elastic,j} \approx -\Delta P_j = k_0 C \partial_{z_i} E_{j \to j+1}, \quad j = 0,1.... \quad (6)$$

$z_i$ is the true tip-sample separation. As a consequence, large signals identify surprisingly soft tips. For our largest signals with $\alpha_{elastic,\,min} \cong -0.65$ the compliance would be $C \cong 2$ m/N which corresponds to the compliance of a contact-mode lever in atomic force microscopy.

It is straightforward to apply the same scheme to the energy of the different vibrational states. Such forces are effective in *I-z* measurements with $V$ = const. corresponding to constant occupation probability of the vibrational states. This requires distinguishing between the externally controlled position of the tip, $z$, and the true tip-sample separation, $z_i$. For small $\delta s$, they fulfill the equation

$$z = C \partial_{z_i} E_j + z_i \quad (7)$$

which can be solved numerically (see supplementary material). The inset of Figure 4(b) displays the numerical derivatives $\partial_{zi} E_0(z_i)$, $\partial_{zi} E_1(z_i)$, and $\partial_{zi} E_{0 \to 1}(z_i) = \partial_{zi} E_1(z_i) - \partial_{zi} E_0(z_i)$. The latter was discussed above. All three are positive for large separation corresponding to attractive forces. Close to the sample all three go negative entailing repulsive forces. In Figure 3(c) the theoretical resonance is plotted versus the externally controlled tip-sample separation, $z$, converting $z_i$ to $z$ according to equation (7) for $j = 1$. In Figure 3(b) the theoretical $\alpha_{elastic}$ is plotted according to equation (6) using $k_0 = 23$ nm$^{-1}$ and $C = 2.1$ m/N. The conformity of the curves is striking. Saturation of $\alpha_{elastic}$ for $\sigma_0 > 1$ μS could be due to a collapse of the lifetime, $\tau_{vib}$, close to the sample. Finally, in Figure 2(a) a measurement of $k$ is displayed for $H_2$:Au(111) at a bias $V \gg V_c$. Thus, the force $-\partial_{zi}E_1$ is effective in that measurement and increases quickly for decreasing $z$. Consequently, $k(z)$ drops off on the left because the true tip-sample separation, $z_i$, lags behind the externally controlled $z$. Converting $z$ to $z_i$, $k(z_i)$ would be constant as would be $k(z)$ for the clean Au(111) in the displayed range of values.

In conclusion, we explained the strong signatures of molecular hydrogen being physisorbed within a tunneling junction by the forces which a single molecule exerts on the confining electrodes in its specific vibrational state. The forces amount up to approximately 30 pN in our experiments. Obviously, even 'well-behaved' and 'well-prepared' tunneling sensors can be unexpectedly soft. However, the experimental arrangement allows control of the vibrational state of a single molecule and detection of the corresponding forces on the confining walls.

**Acknowledgements**

We thank O. Marti and J. Ankerhold for helpful discussions and generous support.

**Authors contribution**

BK designed the experiment and wrote the manuscript. MS and SR conducted the experiments and contributed to manuscript preparation.

**Figures:**

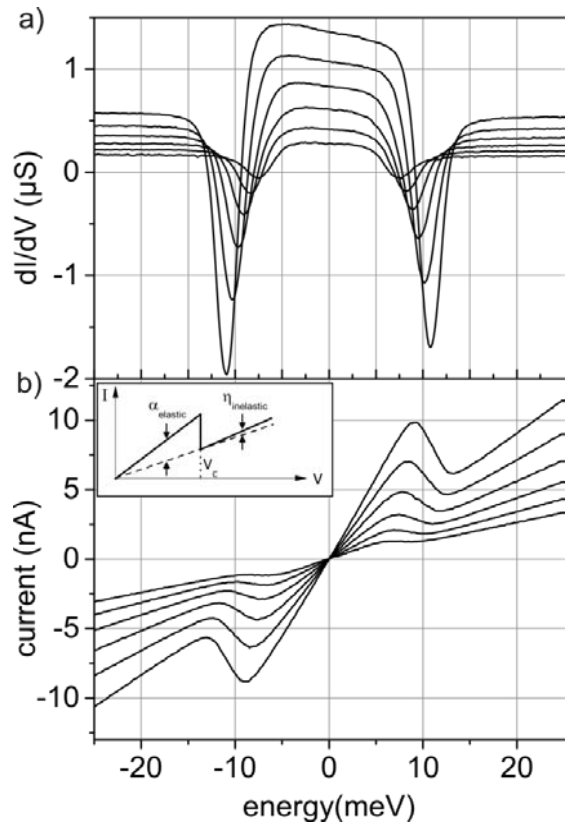

Figure 1: Characteristic curves measured on the system Au(111)-H$_2$-Ag: (a) $\partial_V I$-V curves for different tip-sample separations. (b) Corresponding *I-V* curves showing the characteristic step of the current and the conductivity. They depend strongly on separation. The inset sketches the characteristic parameters: relative change of the elastic conductivity, $\alpha_{elastic}$, and the efficiency of the inelastic channel, $\eta_{inelastic}$.

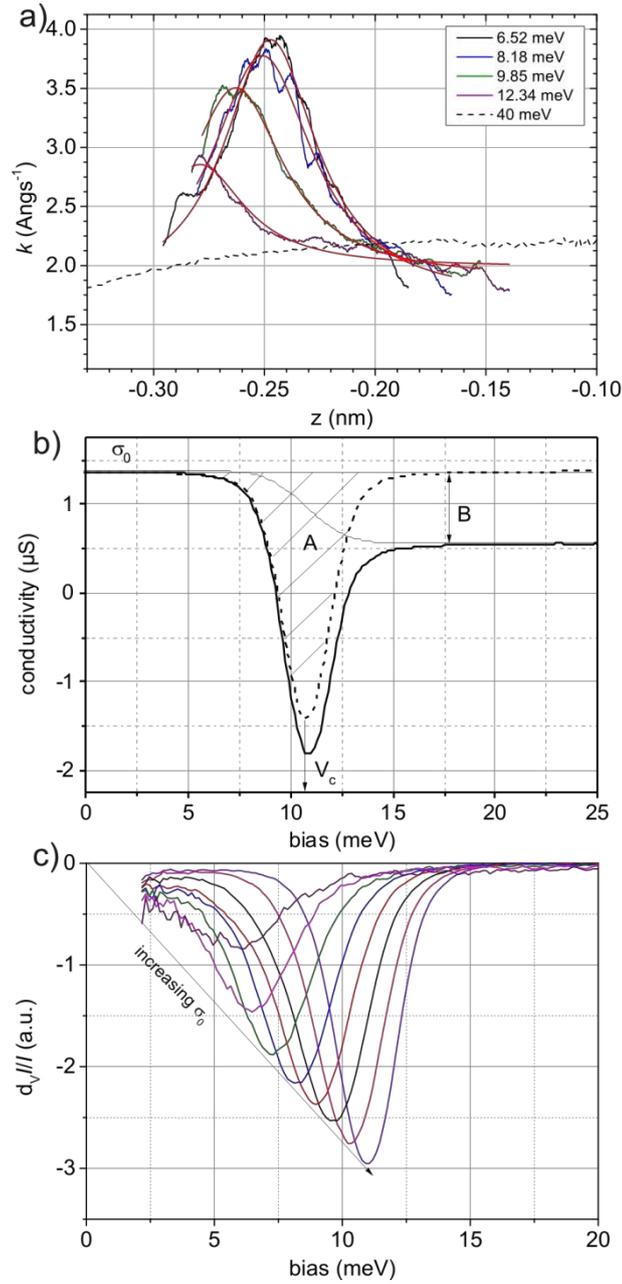

Figure 2: Spectroscopic properties of the system Au(111)-H$_2$-Ag: (a) $k(z;V)$ calculated from consecutive *I-z* measurements at varying bias, *V*. For better legibility, 5 out of 20 curves are shown each being the average of 40 measurements at the same bias. The red solid curves are fits to data using equation (1). The dashed curve is a reference measurement far above resonance. (b) The even part of an exemplary $\partial_V I$-*V* curve together with a fit to data comprising a Gaussian peak, and a step given by an error function centered at the center of the Gaussian peak. The characteristic parameters would be $\alpha_{elastic} = A/(V_c \times \sigma_0)$ and $\eta_{inelastic} = (B-A/V_c)/\sigma_0$. (c) Set of $\partial_V I$-*V* curves normalized to the tunneling current, *I*, for different tip-sample separations (compare equation (4)). The total change of separation is approximately 0.10 nm. The 1/V term has been compensated. The zero-bias conductivity of the purple curve (closest to the sample) is 1.4 µS.

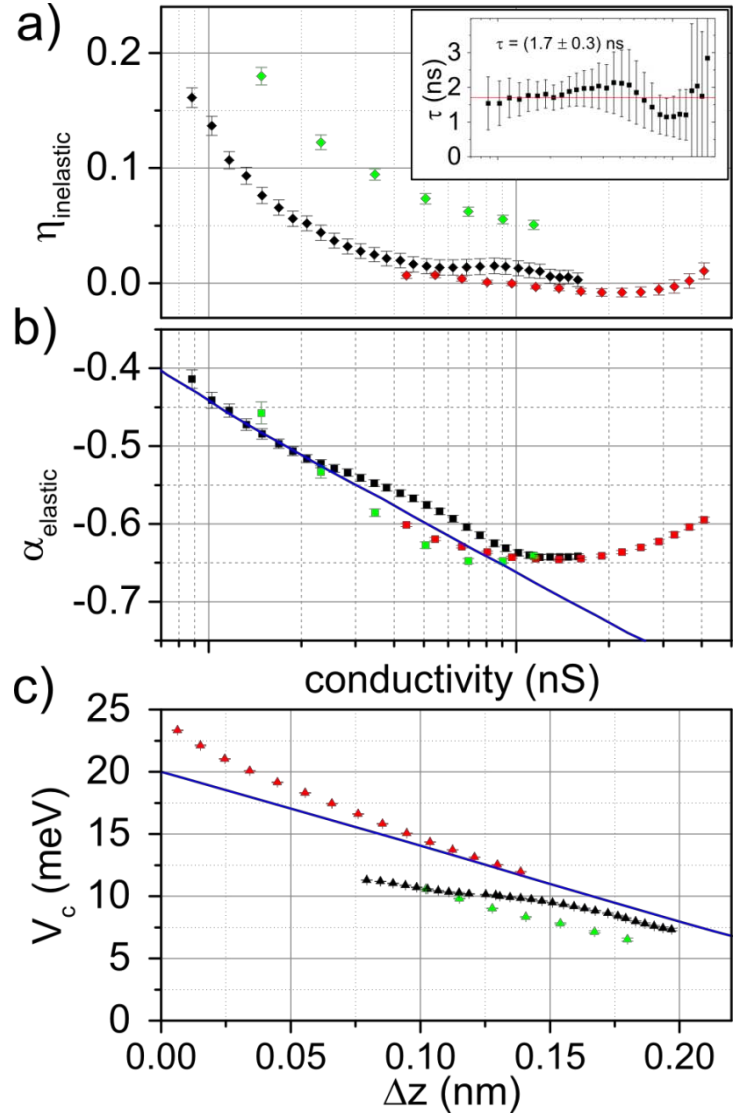

Figure 3: Characteristic parameters derived from fits to I-V curves. The curves in a) and b) are plotted versus the measured zero-bias conductivity, c) is plotted versus separation relative to the set-points before starting the measurement. Parameters are a) the efficiency of the inelastic channel, $\eta_{inelastic}$, b) the relative change of the elastic conductivity at the transition energy, $\alpha_{elastic}$, and c) the peak position in the conductivity, $V_c$, which equals the center of the step. The inset in a) shows the lifetime of the vibration $\tau_{vib} = e/(\sigma_0 V_c \eta_{inelastic})$. The solid blue lines in b) and c) represent the theoretical values of $\alpha_{elastic}$ and $V_c$, respectively.

In b) a conversion of $\sigma_0(z_i)$ is used according to Simmons formula $\sigma_0 = \sigma^* \exp(-k_0 (z_i-z_0))$, $\sigma^* = 6500$ nS, $z_0 = 0.355$ nm; in c) a conversion $z(z_i)$ is used by solving equation (7) with $C = 2.1$ m/N and an offset $\Delta z = z(z_i)-0.355$ nm.

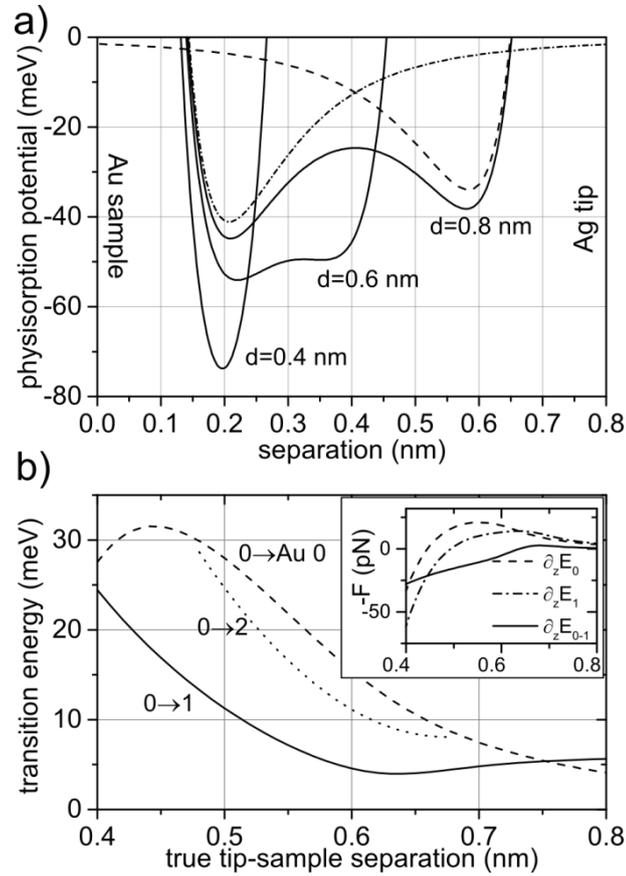

Figure 4: Properties of the physisorption potentials: (a) Physisorption potentials of Au(111) at the left (dash-dotted, minimum on the left) and Ag(111) at the right (dashed, minimum on the right), and the superposition of both at a separation of 0.4 nm, 0.6 nm, and 0.8 nm plotted as indicated vs. separation. (b) Transition energies of the $v = 0 \rightarrow 1$ (solid) and $v = 0 \rightarrow 2$ (dotted) transitions together with the lateral depth of the physisorption well on Au(111) (dashed) due to the presence of the tunneling tip. The inset shows the numerical derivatives of $E_0$, $E_1$, and $E_{0 \rightarrow 1}$ corresponding to the negative forces accompanying the vibrational states '0' and '1' and the transition $v = 0 \rightarrow 1$.